\documentclass[twocolumn,tighten, times]{aastex631}
\usepackage{graphicx}
\usepackage{epstopdf}
\usepackage{float} 
\usepackage{bm}

\usepackage{color}
\usepackage{soul}

\DeclareUnicodeCharacter{02BC}{}

\shorttitle{REST-FRAME NIR DATA}
\shortauthors{Song. ET AL}
\graphicspath{{./}{figures/}}

\begin{document}
\title{Solution to the conflict between the resolved and unresolved galaxy stellar mass estimation from the perspective of JWST}

\author[0000-0002-0846-7591]{Jie Song}
\affil{Deep Space Exploration Laboratory / Department of Astronomy, University of Science and Technology of China, Hefei 230026, People’s Republic of China; \url{xkong@ustc.edu.cn}} 
\affil{School of Astronomy and Space Science, University of Science and Technology of China, Hefei 230026, People’s Republic of China}

\author[0000-0001-9694-2171]{GuanWen Fang}
\affil{Institute of Astronomy and Astrophysics, Anqing Normal University, Anqing 246133, People’s Republic of China; \url{wen@mail.ustc.edu.cn}} 

\author[0000-0001-8078-3428]{Zesen Lin}
\affiliation{Department of Physics, The Chinese University of Hong Kong, Shatin, N.T., Hong Kong S.A.R., People’s Republic of China; \url{zslin@cuhk.edu.hk}}

\author[0000-0003-3196-7938]{Yizhou Gu}
\affil{School of Physics and Astronomy, Shanghai Jiao Tong University, 800 Dongchuan Road, Minhang, Shanghai 200240, People’s Republic of China} 

\author[0000-0002-7660-2273]{Xu Kong}
\affil{Deep Space Exploration Laboratory / Department of Astronomy, University of Science and Technology of China, Hefei 230026, People’s Republic of China; \url{xkong@ustc.edu.cn}} 
\affil{School of Astronomy and Space Science, University of Science and Technology of China, Hefei 230026, People’s Republic of China}

\begin{abstract}
By utilizing the spatially-resolved photometry of galaxies at $0.2<z<3.0$ in the CEERS field, we estimate the resolved and unresolved stellar mass via spectral energy distribution (SED) fitting to study the discrepancy between them. We first compare $M_{\ast}$ derived from photometry with and without the JWST wavelength coverage and find that $M_{\ast}$ can be overestimated by up to 0.2 dex when lacking rest-frame NIR data. The SED fitting process tends to overestimate both stellar age and dust attenuation in the absence of rest-frame NIR data, consequently leading to a larger observed mass-to-light ratio and hence an elevated $M_{\ast}$. With the inclusion of the JWST NIR photometry, we find no significant disparity between the resolved and unresolved stellar mass estimates, providing a plausible solution to the conflict between them out to $z\sim 3$. Further investigation demonstrates that reliable $M_{\ast}$ estimates can be obtained, regardless of whether they are derived from spatially resolved or spatially unresolved photometry, so long as the reddest filter included in the SED fitting has a rest-frame wavelength larger than 10000 \AA.
\end{abstract}


\section{Introduction} \label{sec:intro}

The stellar mass ($M_{\ast}$) is considered as a fundamental physical quantity for describing the properties of a galaxy. Many characteristics of a galaxy, such as its morphology, star formation rate, color, and metallicity, are closely related to $M_{\ast}$ (e.g.,  \citealt{brinchmannPhysicalPropertiesStarforming2004, tremontiOriginMassMetallicity2004a, rodighieroFirstHerschelView2010, bassettCANDELSOBSERVATIONSENVIRONMENTAL2013b, vanderwel3DHSTCANDELSEVOLUTION2014}). Therefore, accurately determining $M_{\ast}$ plays a crucial role in our understanding of galaxy formation and evolution.

Several methods exist for measuring the physical properties of galaxies, with spectral energy distribution (SED) fitting being a particularly important approach \citep{Conroy2013ARAA}. By fitting the SED consisted of multiwavelength photometry, a variety of galaxy properties can be obtained, including $M_{\ast}$, stellar ages, and dust attenuation. However, the accuracy of these properties strongly relies on the available data and the models used in the analysis (e.g., \citealt{Sawicki_1998, marastonEvidenceTPAGB2006, marastonStarFormationRates2010, pforrRecoveringGalaxyStellar2012}). For example, the estimation of $M_{\ast}$ is influenced by the wavelength coverage of the available observations. Previous studies have consistently demonstrated the significance of rest-frame near-infrared (NIR) data in accurately determining $M_{\ast}$ for galaxies (e.g., \citealt{marastonEvidenceTPAGB2006, ilbertGALAXYSTELLARMASS2010a}).
Consequently, to achieve a trustworthy $M_{\ast}$, it is essential to incorporate observed-frame NIR data for galaxies at low and intermediate redshifts and mid-infrared data for galaxies at higher redshifts. 


Recent advancements in high-resolution images (e.g., CANDELS; \citealt{groginCANDELSCOSMICASSEMBLY2011, koekemoerCANDELSCOSMICASSEMBLY2011}) allow us to study the spatial distribution of stellar matter within galaxies by spatially resolved (e.g., pixel-by-pixel) SED fitting at high redshifts. Interestingly, some previous works reported significant differences between the resolved (via pixel-by-pixel SED fitting, $M_{\rm \ast, resolved}$) and unresolved (derived from the integrated flux of the galaxy, $M_{\rm \ast, unresolved}$) stellar mass estimations. With 67 nearby galaxies from Sloan Digital Sky Survey (SDSS; \citealt{eisensteinSDSSIIIMASSIVESPECTROSCOPIC2011}), \cite{sorbaMissingStellarMass2015} found that $M_{\rm \ast, resolved}$ derived based on u, g, r, i, z, and NUV data was approximately 13\% (0.06 dex) larger than $M_{\rm \ast, unresolved}$. This deviation increases gradually with specific star formation rate (sSFR) and reaches a maximum of 25\% (0.12 dex) at an sSFR of $10^{-8}yr^{-1}$.
Similarly, \cite{sorbaSpatiallyUnresolvedSED2018} studied a high-redshift galaxy sample from the Hubble eXtreme Deep Field \citep{illingworthHSTEXTREMEDEEP2013} and found that the difference between $M_{\rm \ast, resolved}$ and $M_{\rm \ast, unresolved}$ is small for galaxies with small sSFR (predominantly consist of low-redshift galaxies) but increases rapidly when sSFR $\gtrsim 10^{-9.5}yr^{-1}$.

\begin{figure*}[htb!]
\centering
\includegraphics[width=0.92\textwidth, height=0.48\textwidth]{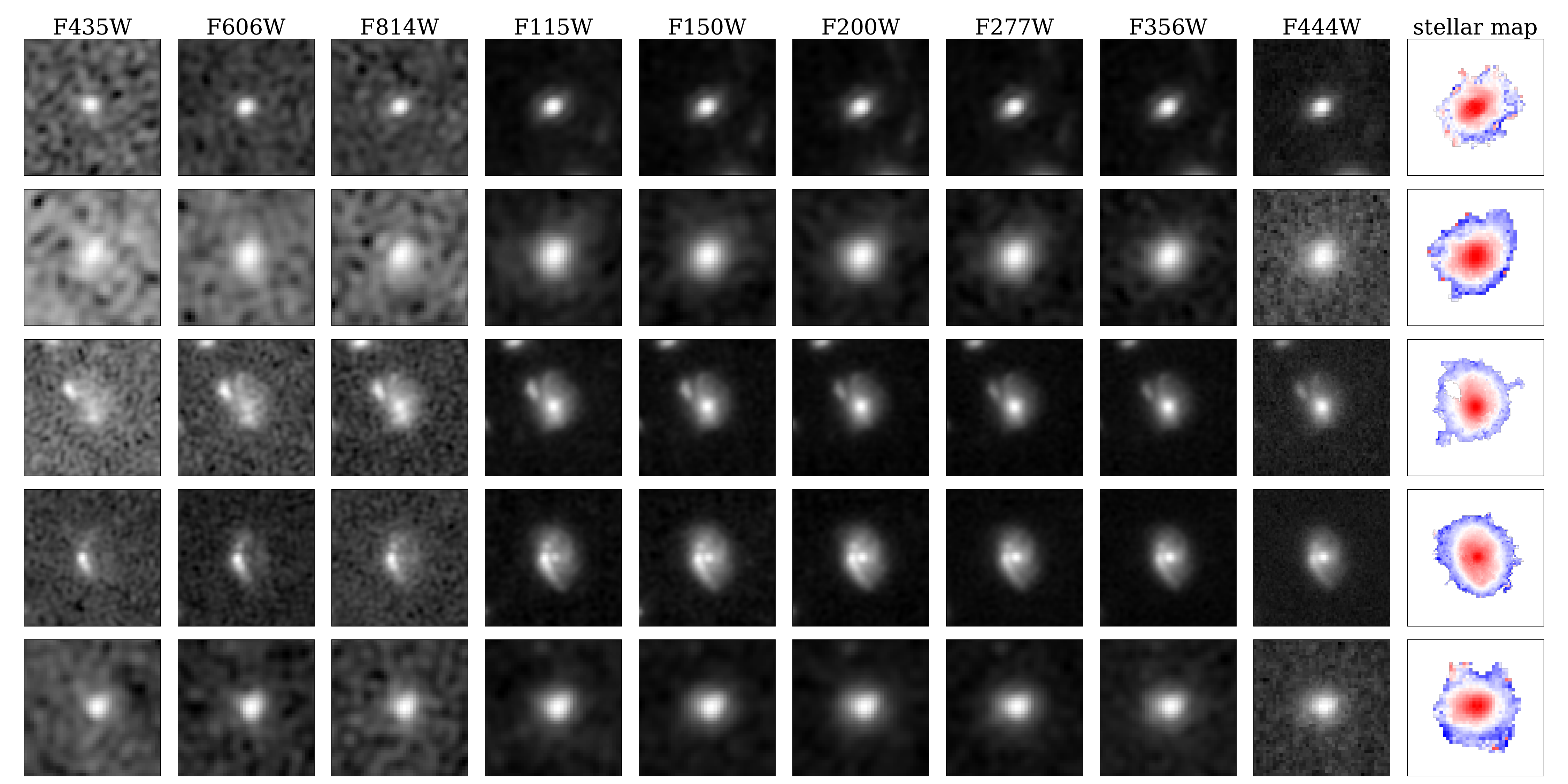}
\caption{Images from different bands and the estimated stellar mass maps of a few example galaxies. The stellar mass maps are derived using the pixel-by-pixel SED fitting method.}
\label{fig1}
\end{figure*}

Several studies have attributed the difference between $M_{\rm \ast, resolved}$ and $M_{\rm \ast, unresolved}$ to the phenomenon of outshining. Young and massive stars possess lower $M_{\ast}/L$ ratios but are  orders of magnitude brighter compared to older stellar populations. 
When considering the integrated flux of a galaxy, the optical luminosity is dominated by these young stars. Therefore, in SED fitting, the model tends to prioritize fitting the $M_{\ast}/L$ of the young stellar population, potentially ignoring the mass contribution of old stellar populations (e.g., \citealt{Sawicki_1998,  papovichStellarPopulationsEvolution2001, marastonStarFormationRates2010, gimenez-arteagaSpatiallyResolvedProperties2022a}).

But it is important to highlight that, as aforementioned, the availability of rest-frame NIR data significantly influences the results of SED fitting. Notably, the longest observed wavelength of HST is approximately 16000 \AA\ and falls short of capturing the rest-frame NIR emission for high-redshift galaxies. Consequently, some previous studies, particularly those involving high-redshift samples, did not have rest-frame NIR data in the pixel-by-pixel SED fitting when they used HST data to ensure high spatial resolution. This makes it difficult to accurately determine whether the disparity between $M_{\rm \ast, resolved}$ and $M_{\rm \ast, unresolved}$ arises from the lack of rest-frame NIR data or the outshining effect. Fortunately, JWST offers a promising solution by providing high-resolution rest-frame NIR imagining data even up to $z\sim 3$. 
Leveraging the capabilities of the JWST, in this work, we utilize images obtained from the Cosmic Evolution Early Release Science (CEERS) program (PI: Finkelstein) to conduct a comparative study between resolved and unresolved $M_{\ast}$ estimated via SED fitting with a larger coverage of the rest-frame wavelength, notably encompassing rest-frame NIR observations at high redshift. 
Our result shows there is no significant difference between $M_{\rm \ast, resolved}$ and $M_{\rm \ast, unresolved}$ after considering the data from JWST, 
which means that we can obtain a reliable $M_{\ast}$ estimation with spatially unresolved photometry when the wavelength coverage is large enough, without the need to consider the outshining effect (if any).


The layout of this paper is as follows. In Section \ref{sec:data}, we introduce the data we used and the method to estimate $M_{\rm \ast, resolved}$ and $M_{\rm \ast, unresolved}$. The main results are shown in Section \ref{sec:result}. We give a brief discussion in Section \ref{sec:dis} and conclude in Section \ref{sec:sum}. Throughout the paper, we adopt a flat $\Lambda$CDM cosmology with $H_0 = 70~\mathrm{km~s^{-1}~Mpc^{-1}}$, $\Omega_{\rm{m}}=0.3$, and $\Omega_{\rm{\Lambda}}=0.7$, and a \cite{chabrierGalacticStellarSubstellar2003} initial mass function (IMF).

\section{Data and Method} \label{sec:data}

\subsection{Sample selection}

The CEERS program (Finkelstein et al., in prep.) is one of the early release science surveys of JWST. The mosaicked images and weight maps utilized in this study were processed by \cite{valentinoAtlasColorselectedQuiescent2023}, based on the public grizli software package \citep{2022zndo...6985519B}. The HST images in the same field were also processed by  \cite{valentinoAtlasColorselectedQuiescent2023}. For this investigation,  we employ imaging mosaics obtained from JWST in six broad-band filters (F115W, F150W, F200W, F277W, F356W, and F444W). Due to the proximity of the central wavelengths of filters JWST/F115W and HST/F125W, and filters JWST/F150W and HST/F160W, we do not include these two HST filters in this study, namely, only the F435W, F606W, and F814W filters from HST are utilized.

Our galaxy sample is constructed based on the 3D-HST catalog (\citealt{brammer3DHSTWIDEFIELDGRISM2012, skelton3DHSTWFC3SELECTEDPHOTOMETRIC2014, momcheva3DHSTSURVEYHUBBLE2016}). They fit galaxy SEDs in the wavelength range of $0.3–8.0~\mu$m to estimate photometric redshift and other galaxy physical properties. For a more detailed description, we refer readers to their papers. Our sample comprises galaxies within the CEERS field that are bright and massive enough
to facilitate reliable SED fitting. We require the result from the 3D-HST catalog to be reliable by setting $\rm use\_phot = 1$ and $\rm flags\leq 2$ and restrict our sample to galaxies with $\log(M_{\ast}/M_{\odot}) > 9$ at $0.2 < z < 3.0$. Additionally, we only include galaxies that are detected ($\rm S/N > 3$) in all six JWST filters and in at least two HST filters. Sources at the boundary of the field are removed. Finally, 1060 galaxies are contained in our sample.

\subsection{Analysising method} \label{sec:method}

In this section, we provide a brief overview of the method employed to construct the stellar mass maps in our study.
To begin, we adopt the PSF models from grizli-psf-library provided by Brammer et al. (in prep) and create convolution kernels with the {\tt\string photutils} package \citep{2022zndo...6825092B} by optimizing the measurements of kernel performance proposed by \cite{anianoCommonResolutionConvolutionKernels2011}. The shorter-wavelength images are then PSF-matched to the F444W band since the PSF FWHM of this band is the largest.
Subsequently, the final images used to generate the spatially-resolved stellar mass maps are resampled to a common pixel scale of $0\farcs04$. The images used in this work are also corrected for the Milky Way extinction \citep{schlaflyMEASURINGREDDENINGSLOAN2011} assuming the \cite{fitzpatrickCorrectingEffectsInterstellar1999} extinction curve with $R_V = 3.1$. 

For each galaxy in our sample, we generate a cutout that is 1.5 times larger than the segmentation map provided by \cite{valentinoAtlasColorselectedQuiescent2023}. Within this cutout, we perform SED fitting on all pixels to obtain the stellar mass map of the galaxy. We fix the redshift of all pixels to $z_{\rm best}$ retrieved from the 3D-HST catalog and utilize the {\tt\string CIGALE} program \citep{boquienCIGALEPythonCode2019} to estimate the $M_{\ast}$ of each pixel. In brief, we adopt the delayed-$\tau$ star formation history, \cite{bruzualStellarPopulationSynthesis2003} stellar population models, \cite{inoueRestframeUltraviolettoopticalSpectral2011} nebular emission-line models, and \cite{charlotSimpleModelAbsorption2000} dust attenuation law in the fitting. In Figure \ref{fig1}, we show the multi-band images and the corresponding stellar mass maps of a few example galaxies.


The SED fitting result can be influenced by many factors, such as the treatment of the thermally pulsating asymptotic giant branch stage, the choice of IMF, and assumptions regarding the star formation history (e.g., \citealt{Sawicki_1998, marastonEvidenceTPAGB2006, marastonStarFormationRates2010, pforrRecoveringGalaxyStellar2012}), leading to large model-dependent systematic uncertainties in estimating $M_{\ast}$. However, the aim of this study is to study the difference between $M_{\rm \ast,\ resolved}$ and $M_{\rm \ast,\ unresolved}$ under the same model assumptions.
Therefore, the systematic bias due to the model selection will not be considered in this study. Fortunately, once the redshift is well-determined, $M_{\ast}$ is one of the most robust parameters estimated from the SED-fitting, while other parameters are generally more difficult to estimate because of the degeneracy (e.g., \citealt{caputiSPITZERBRIGHTULTRAVISTA2015, pacificiArtMeasuringPhysical2023}). Thanks to the richness of multi-band observations, the uncertainty of the photometric redshifts in the 3D-HST catalog is $\sigma_{\Delta z/(1+z)}\sim0.02$, which is small enough to ensure reliable $M_{\ast}$ estimations. Additionally, within our sample, we have 82 sources with spectroscopic redshifts. In the subsequent analysis, we also independently verify the results using the subset of sources with spectroscopic redshifts and find that they align with the results obtained from the entire sample.

\begin{figure}[htb!]
\centering
\includegraphics[width=0.48\textwidth, height=0.65\textwidth]{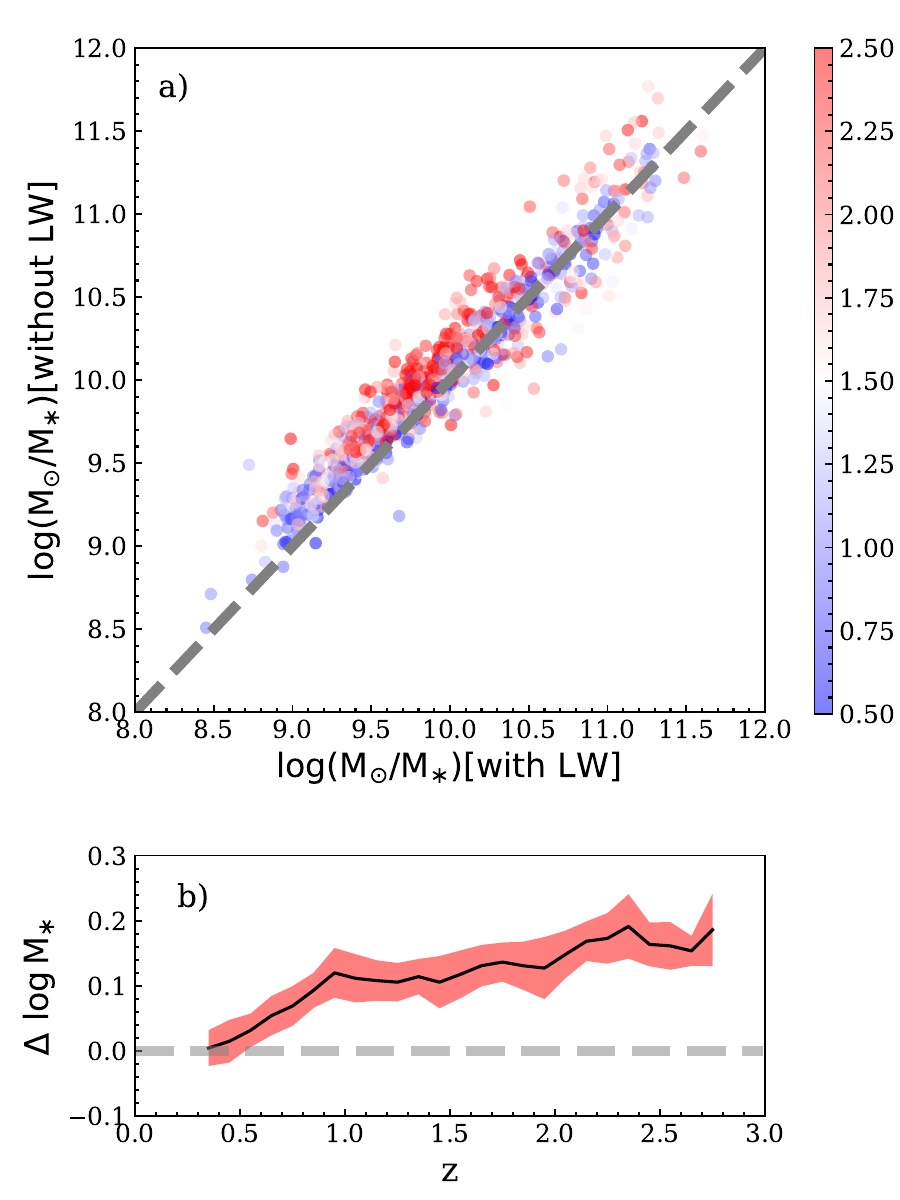}
\caption{Upper panel: a direct comparison between $M_{\ast}$ estimated with and without LW bands color-coded by redshift. The gray dashed line is the one-to-one relation. Lower panel: the difference between them (i.e., $\log (M_{\rm \ast, without~LW}/M_{\rm \ast, with~LW})$) as a function of redshift. The black solid line and red-shaded area represent the median and 1-$\sigma$ uncertainty, respectively.}
\label{fig2}
\end{figure}

\begin{figure*}[htb!]
\centering
\includegraphics[width=0.9\textwidth, height=0.45\textwidth]{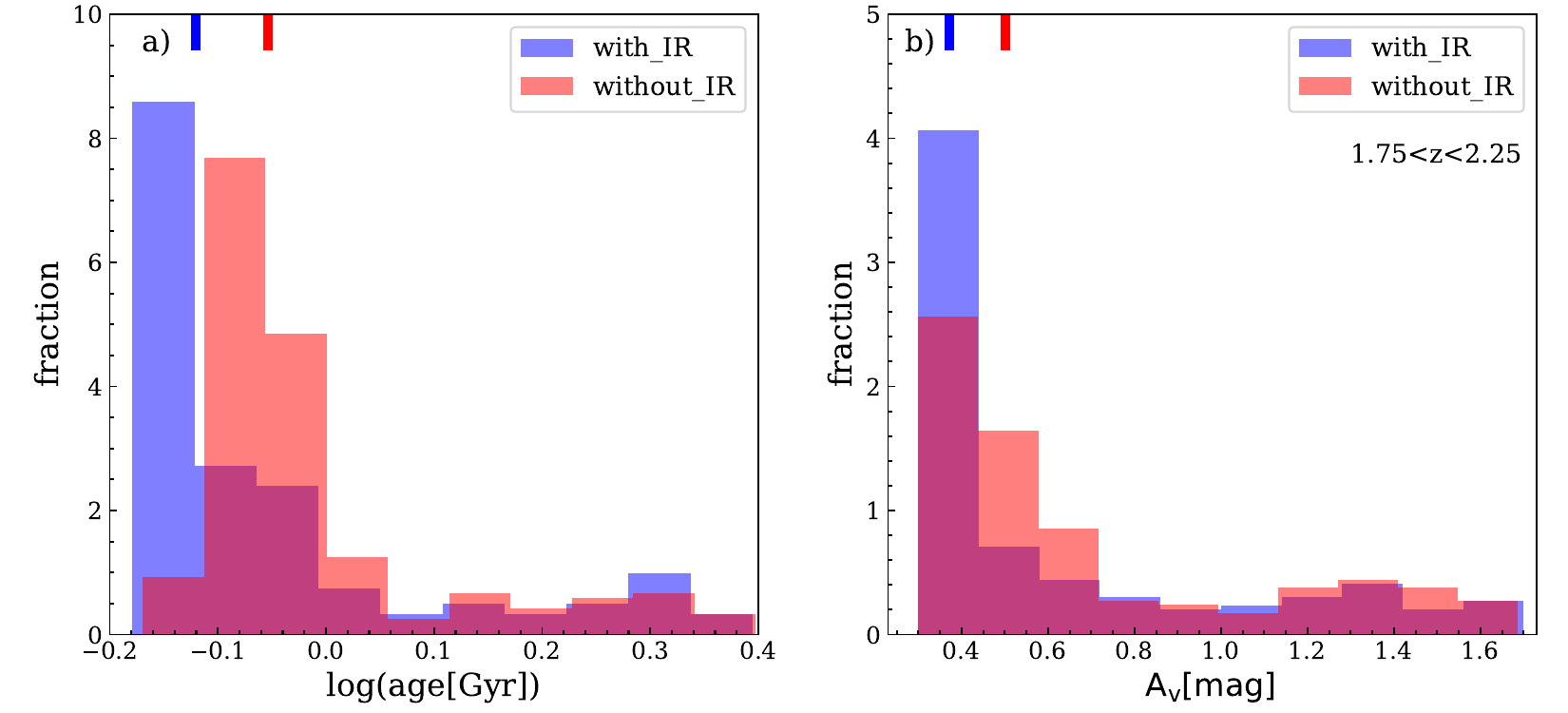}
\caption{Distributions of stellar age (left panel) and $A_{\rm V}$ (right panel) for galaxies at $z\sim 2$, with blue and red colors representing the distributions from the fittings with and without LW bands, respectively. The median values are denoted as bars at the top of the figure. The stellar age and $A_{\rm V}$ are indeed overestimated when rest-frame NIR data is not included in the SED fitting.}
\label{fig3}
\end{figure*}

\section{Results} \label{sec:result}

\subsection{Comparisons between fittings with and without rest-frame NIR data}
\label{subsec:comp_nir}

Before investigating the disparity between $M_{\rm \ast, resolved}$ and $M_{\rm \ast, unresolved}$, we conduct an examination of the influence of the rest-frame NIR data on the $M_{\ast}$ estimation. The total flux catalog extracted by \cite{valentinoAtlasColorselectedQuiescent2023} is used to estimate $M_{\ast}$ here via the {\tt\string CIGALE} program with the same configuration described in Section \ref{sec:method}. For simplicity, $M_{\ast}$ derived from the fittings using photometry with and without the long wavelength (LW) bands (i.e., F200W, F277W, F356W, and F444W) are denoted as $M_{\rm \ast,with~LW}$ and $M_{\rm \ast, without~LW}$, respectively. The comparison between them is presented in Figure \ref{fig2}.

In the upper panel of Figure \ref{fig2}, we present the distribution of $M_{\rm \ast, without~LW}$ and $M_{\rm \ast, with~LW}$ for our sample color-codded by redshift. It is evident that $M_{\rm \ast, without~LW}$ tends to be larger than $M_{\rm \ast,with~LW}$. To further investigate this issue, we examine the difference between them (i.e., $\log (M_{\rm \ast, without~LW}/M_{\rm \ast, with~LW})$) as a function of redshift in the lower panel. The solid black line represents the median difference, while the red-shaded region represents the 1-$\sigma$ uncertainty estimated by the bootstrapping method. It can be seen that there is no significant difference between the two at $z<0.5$.
However, at higher redshift, a notable disparity emerges, reaching up to 0.2 dex. We attribute this redshift trend to the fact that the longest bands involved in the fitting without LW bands (i.e., F150W) can not trace the rest-frame NIR emission ($\gtrsim 10000$ \AA ) anymore at $z\gtrsim 0.5$.
Interestingly, a similar trend was reported by \cite{ilbertGALAXYSTELLARMASS2010a} in which the authors compared $M_{\ast}$ estimated with and without IRAC data, akin to our comparison with and without LW bands. 
In their work, at $z<1.5$, the IRAC data have a negligible impact on the $M_{\ast}$ estimation, since the K band (the longest band available except the IRAC data) can still cover the rest-frame NIR data. However, at $z > 1.5$, lack of IRAC data could lead to an overestimation of $M_{\ast}$ up to 0.1 dex since the K band only covers rest-frame $\sim 9000$ \AA\ in this redshift range.

Moreover, we have further validated this result using simulated data. Using the CIGALE program, we construct a sophisticated Stellar Population Synthesis (SPS) library and randomly select a sample of 1500 SEDs from our library. The model employed in this simulation is the same as the one mentioned in Section \ref{sec:method}. By fixing the redshift to 2, we are able to scrutinize the ramifications of incorporating rest-frame NIR data, employing the same filters with the observation. Recognizing the inherent variations in flux error medians across different filters in real observations, the standard deviation ($\sigma$) of the Gaussian noise has been tailored to match the median flux error of each band. Subsequently, we have proceeded to derive stellar mass from the perturbed fluxes, leveraging the capabilities of the CIGALE software. With this simulated data, the median deviation between the estimated and true stellar mass is approximately 0.08 dex in the absence of rest-frame NIR data, while this discrepancy reduces to a mere 0.02 dex when incorporating rest-frame NIR data. Furthermore, the dispersion in mass estimates is diminished when rest-frame NIR data is utilized. Specifically, the standard deviation amounts to 0.04 dex with NIR inclusion, compared to 0.14 dex when NIR is absent in the fitting. This indicates that without the rest-frame NIR data, the measurement of $M_{\ast}$ is indeed biased statistically.

There are several potential factors that could contribute to an overestimation of $M_{\ast}$ in the absence of the rest-frame NIR data. Since $M_{\ast}$ is determined by multiplying the galaxy's luminosity by the $M_{\ast}/L$ ratio, the inclusion or exclusion of rest-frame NIR data may impact the derived $M_{\ast}/L$ ratio. Numerous studies have demonstrated that factors such as stellar age and dust attenuation can influence the observed $M_{\ast}/L$ ratio (e.g., \citealt{lejaUVJMoreEfficient2019, millerEarlyJWSTImaging2022}). Using galaxies at $z\sim 2$, we compare the distributions of stellar age and dust attenuation ($A_{\rm V}$) obtained from the fittings with and without LW bands and present them in Figure \ref{fig3}. Evidently, both the stellar age and $A_{\rm V}$ tend to be overestimated when performing SED fitting without rest-frame NIR data. The median stellar age and $A_{\rm V}$ are 0.76 (0.88) Gyr and 0.37 (0.50) mag, respectively, for the results with (without) LW bands. To assess the statistical significance of the differences between them, we apply the Kolmogorov-Smirnov test and find that the $p$-values for stellar age and $A_{\rm V}$ are both smaller than $10^{-3}$, suggesting substantial differences between the two distributions.
We thus conclude that the stellar age and $A_{\rm V}$ are indeed overestimated when the rest-frame NIR data is not included in the SED fitting. For a given luminosity, an older stellar population and a large $A_{\rm V}$ can lead to a larger observed $M_{\ast}/L$ and consequently an overestimation of $M_{\ast}$. The physical reason for these differences when lacking rest-frame observations at wavelength $\gtrsim 10000$ \AA\ deserves further investigation.

\subsection{Comparison between resolved and unresolved stellar mass}

\begin{figure}[htb!]
\centering
\includegraphics[width=0.48\textwidth, height=0.65\textwidth]{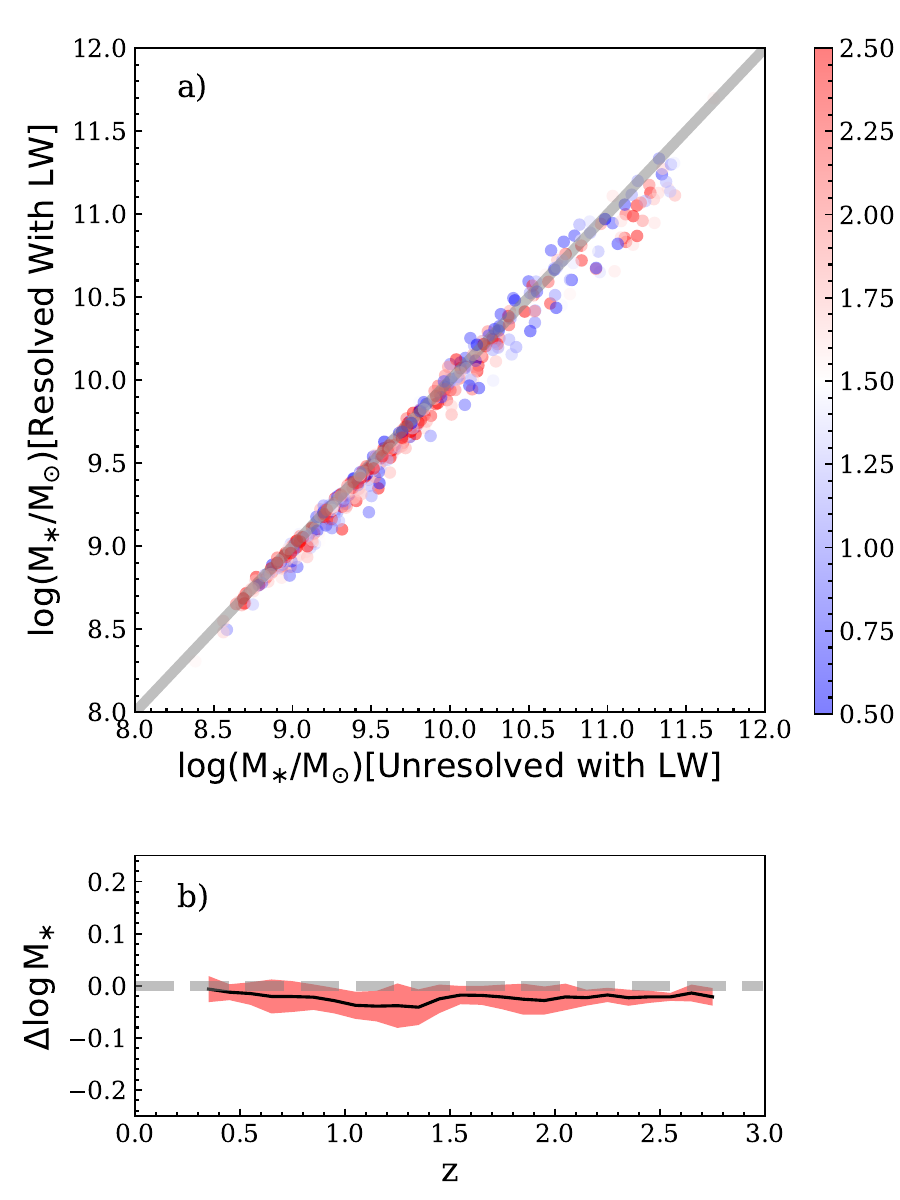}
\caption{Similar to Figure \ref{fig2} but for the comparison between $M_{\rm\ast, resolved}$ and $M_{\rm\ast, unresolved}$.
It is evident that there is no significant difference between $M_{\rm\ast, resolved}$ and $M_{\rm\ast, unresolved}$. }
\label{fig4}
\end{figure}

\begin{figure*}[htb!]
\centering
\includegraphics[width=\textwidth]{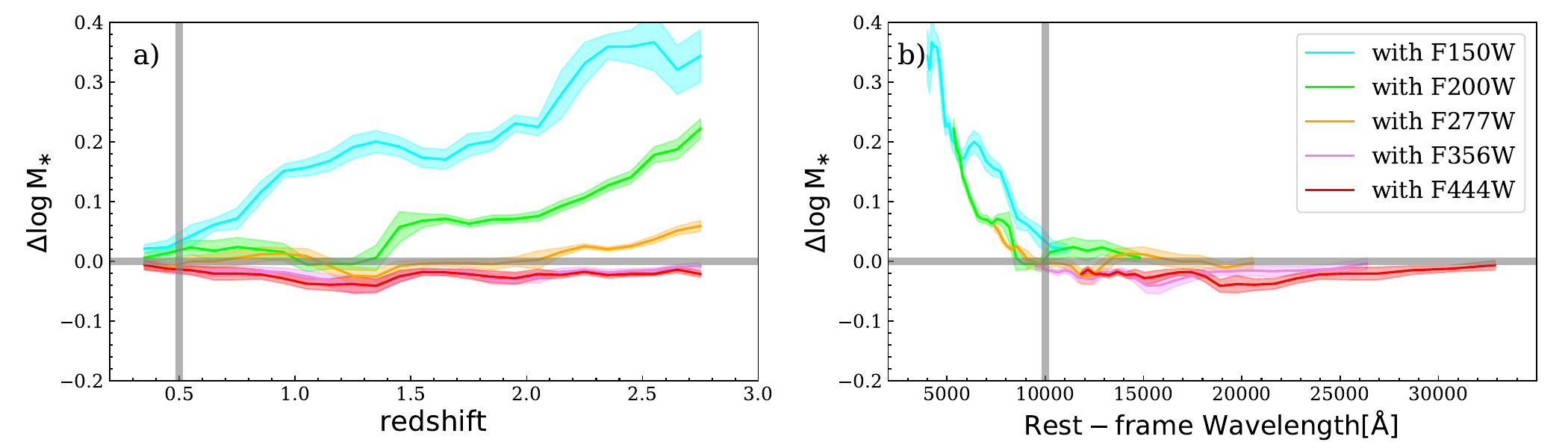}
\caption{Difference between $M_{\rm \ast, resolved}$ and $M_{\rm \ast, unresolved}$ as a function of redshift (left panel) and the reddest rest-frame wavelength included in the SED fitting for $M_{\rm \ast, resolved}$ estimate (right panel). $M_{\rm \ast, unresolved}$ is estimated using all available data (i.e., from F435W to F444W). The cyan, lime, orange, violet, and red curves represent the median differences between $M_{\rm \ast, resolved}$ and $M_{\rm \ast, unresolved}$ when the reddest filters used are F150W, F200W, F277W, F356W, and F444W for $M_{\rm \ast, resolved}$ estimates, respectively, while the color-shaded regions are the corresponding 1-$\sigma$ uncertainties. The vertical grey line in the left panel indicates $z=0.5$ below which there is no significant difference between $M_{\rm \ast, resolved}$ and $M_{\rm \ast, unresolved}$, while the similar line in the right panel locates rest-frame wavelength of 10000 \AA\ beyond which the differences in $M_\ast$ are negligible.}
\label{fig5}
\end{figure*}

Many previous studies showed that $M_{\rm\ast,unresolved}$ is underestimated compared to $M_{\rm\ast,resolved}$, particularly at high redshift (e.g., \citealt{zibettiResolvedStellarMass2009,sorbaSpatiallyUnresolvedSED2018, moslehGalaxySizesPerspective2020a}). However, as aforementioned, obtaining $M_{\rm\ast, resolved}$ measurements relies on high spatial resolution images, primarily obtained from HST. It is important to note that the reddest filter of HST is F160W, which can only trace the rest-frame optical ($\lesssim 8000$ \AA) emission at $z>1$. Consequently, relying solely on these observations may lead to an overestimation of $M_{\ast}$ for each pixel (see Section \ref{subsec:comp_nir}).

To address this issue and mitigate the potential bias in $M_{\ast}$ estimation, it is crucial to include rest-frame NIR observations, such as those provided by JWST. Here, we use PSF-matched HST and JWST images (from F435W to F444W) to further investigate the difference between $M_{\rm \ast, resolved}$ and $M_{\rm \ast, unresolved}$. Before deriving $M_{\rm \ast, resolved}$, we need to account for the potential bias introduced by the poor signal-to-noise ratio (S/N) of individual pixels during the SED fitting, as discussed in \cite{gallazziSTELLARMASSTOLIGHTRATIOS2009}. To this end, we only consider pixels with S/N larger than 3 in all six JWST bands. 
Regarding the HST data in the CEERS field, we find that the S/N, particularly in the F435W band, is generally much lower compared to the JWST images. 
Applying a similar S/N threshold to the HST images leads to a significant reduction of the sample size. To obtain a galaxy sample with a reasonable size, we determined not to impose the S/N restrictions on the HST images. However, even if such restrictions are applied, we still obtain almost unchanged conclusions.
Nonetheless, the same as \cite{sorbaSpatiallyUnresolvedSED2018}, the pixel number included in the analysis for each galaxy is required to be $\geq 64$ since the uncertainty in mass estimation becomes inordinately large below this threshold. Finally, 535 galaxies in total are included in the following analysis. To estimate $M_{\rm \ast, resolved}$, we sum the stellar masses of all pixels within the segmentation map that satisfy the aforementioned criteria. Then we estimate $M_{\rm \ast, unresolved}$ by summing the fluxes from all the corresponding pixels and performing SED fitting with the same parameter configuration.

The comparison between the resolved and unresolved stellar mass is depicted in Figure \ref{fig4}.
Notably, after incorporating JWST data, there is no significant difference observed between the resolved and unresolved stellar masses. This result is further emphasized in the lower panel of Figure \ref{fig4}, in which we plot the difference between $M_{\rm\ast, resolved}$ and $M_{\rm\ast, unresolved}$ (i.e., $\log (M_{\rm\ast, resolved}/M_{\rm\ast, unresolved})$) as a function of redshift.
It is evident that, when accounting for the associated errors, $M_{\rm\ast, resolved}$ and $M_{\rm\ast, unresolved}$ agree well, while the median bias and its 1-$\sigma$ uncertainty are all about 0.02 dex.
We calculate the mass-weighted age by weighting the stellar ages of all selected pixels by their stellar masses within one galaxy and find that the median mass-weighted age of the sample from the resolved fitting (0.83 Gyr) is slightly smaller than the one obtained from the unresolved photometry (0.97 Gyr), which is consistent with the tiny difference that $M_{\rm \ast, resolved}$ is slightly smaller than $M_{\rm \ast, unresolved}$. Note that outshining would expect a larger resolved mass-weighted age compared to the unresolved one. Although outshining might have an effect on the $M_{\ast}$ estimation when the rest-frame NIR data is not included, our result suggests that incorporating the rest-frame NIR data in the fitting could correct for the potential impact of outshining (if any) and return a reliable measurement of $M_{\ast}$, regardless of whether spatially resolved or unresolved photometry is used.
Different SED fitting codes may employ different data handling methods. To validate our findings, we conduct similar analyses using the {\tt\string FAST} \citep{kriekULTRADEEPNEARINFRAREDSPECTRUM2009} and {\tt\string BAGPIPES} \citep{carnallInferringStarFormation2018} programs. Remarkably, the results obtained from these alternative programs still align with those obtained from the {\tt\string CIGALE} fitting. By incorporating the JWST data, we consistently find no significant difference between the resolved and unresolved stellar masses. Moreover, different stellar populations may also produce notable differences in stellar mass estimates. We have also estimated both resolved and unresolved stellar masses for our sample, employing the CIGALE program, with the stellar population shifting from \cite{bruzualStellarPopulationSynthesis2003} to \cite{marastonEvolutionaryPopulationSynthesis2005}. With this new model, there is also no significant difference between $M_{\ast,\rm resolved}$ and $M_{\ast,\rm unresolved}$. Further efforts are still needed to examine the consistency between $M_{\ast,\rm resolved}$ and $M_{\ast,\rm unresolved}$ estimations with other stellar population models.

\section{Discussion} \label{sec:dis}

In many previous studies, $M_{\rm \ast, resolved}$ is found to be larger than $M_{\rm \ast, unresolved}$.
However, with only HST images considered, \cite{wuytsSMOOTHERSTELLAR2012} found that there is no significant difference between $M_{\rm \ast, resolved}$ and $M_{\rm \ast, unresolved}$ at $1.5<z<2.5$ when they used the Voronoi two-dimensional binning technique to ensure that the minimum S/N of each bin in the $H_{\rm 160}$ image was 10. Interestingly, the authors found that when directly measuring $M_{\rm \ast, resolved}$ without binning, the difference between $M_{\rm \ast, resolved}$ and $M_{\rm \ast, unresolved}$ for star-forming galaxies at $z\sim2.0$ could be as large as 0.2 dex. They attributed this discrepancy to the presence of pixels on the outskirts of galaxies that may lack sufficient color constraints, leading to an overestimation of stellar mass when pixel binning is not considered. In our study, pixels with low S/N on the outskirts of galaxies are ignored. Even so, our additional check still confirms that when the LW data was not considered, $M_{\rm \ast, resolved}$ is larger than $M_{\rm \ast, unresolved}$, at least at high redshift, which implies that our data does not support the high-S/N explanation for the consistency between $M_{\rm \ast, resolved}$ and $M_{\rm \ast, unresolved}$.
On the other hand, \cite{sorbaMissingStellarMass2015} studied the effect of image resolution on the difference between $M_{\rm \ast, resolved}$ and $M_{\rm \ast, unresolved}$ and found that the discrepancy diminishes when the physical scale of individual pixels is larger than 3 kpc, which is larger than the typical size of star-forming regions \citep{gusevParametersBrightestStar2014}. They speculated that the disappearance of the difference observed in \cite{wuytsSMOOTHERSTELLAR2012} may be attributed to the mixing of young and old stars when employing pixel binning, reducing the bias caused by the outshining effect.


Since the inclusion of rest-frame NIR data is crucial for the accurate estimation of $M_{\ast}$, determining the minimum requirement of the reddest filter that should be included in SED fitting is important. In Figure \ref{fig2}, we can see that without adding the LW data, we cannot obtain a reliable $M_{\ast}$ estimate for galaxies at $z>0.5$. Consequently, it seems that observations with a rest-frame wavelength longer than about 1$\mu$m (i.e., the regime covered by F160W at $z\sim0.5$) are essential for obtaining accurate $M_{\ast}$ estimates. We here further investigate this issue by gradually decreasing the longest wavelength when estimating $M_{\rm \ast, resolved}$ and study the difference between the corresponding $M_{\rm \ast, resolved}$ and $M_{\rm \ast, unresolved}$. Note that $M_{\rm \ast, unresolved}$ used here is the one derived from the fittings with all available photometry (i.e., with the LW data). The results are shown in Figure \ref{fig5} where curves in different colors (cyan, lime, orange, violet, and red) represent the difference between $M_{\rm \ast, resolved}$ and $M_{\rm \ast, unresolved}$ when the reddest filters used are F150W, F200W, F277W, F356W, and F444W for $M_{\rm \ast, resolved}$ estimates, respectively. Panel (a) shows that we can obtain a reliable $M_{\rm \ast, resolved}$ at $z\lesssim0.5$, $z\lesssim1.4$, $z\lesssim2.1$, $z\lesssim2.8$, and $z\lesssim2.8$ when the reddest filters used are F150W, F200W, F277W, F356W, and F444W, respectively. We also show the difference as a function of the reddest rest-frame wavelength included in the SED fitting in Panel (b). There is no significant difference between the resolved and unresolved $M_\ast$ when the longest wavelength employed in SED fitting extends beyond the rest-frame 10000 \AA, which is consistent with the results discussed in Section \ref{subsec:comp_nir}.
Therefore, we argue that an unbiased $M_{\ast}$ estimation requires the reddest filter included in the SED fitting to have a rest-frame wavelength larger than 10000 \AA.

\section{Summary} \label{sec:sum}

Utilizing the high-resolution data from JWST and HST, we estimate $M_{\rm \ast, resolved}$ and $M_{\rm \ast, unresolved}$ using the spatially resolved and spatially unresolved photometry of galaxies with $\log(M_{\ast}/M_{\odot}) >9$ at $0.2<z<3.0$ in the CEERS field. Our main conclusions are as follows.

(1) The inclusion of rest-frame NIR data is crucial for accurate measurement of galaxy $M_{\ast}$. In the absence of rest-frame NIR data, the SED fitting process tends to yield higher stellar ages and dust attenuations, leading to an overestimation (approximately $0.1\sim 0.2$ dex) of $M_{\ast}$.

(2) After incorporating the data from JWST, there is almost no difference between $M_{\rm \ast, resolved}$ and $M_{\rm \ast, unresolved}$, suggesting that we could correct for the potential impact of outshining (if any) and obtain a reliable measurement of $M_{\ast}$.

(3) When the reddest filter included in the SED fitting has a rest-frame wavelength larger than 10000 \AA, both resolved and unresolved photometry can yield consistent and nearly unbiased measurements of stellar mass.

Benefiting from the JWST photometry, we provide a plausible solution to the conflict between the resolved and unresolved $M_\ast$ estimation. This solution works out to $z\sim 3$ and will be further examined for galaxies at $z\gtrsim 3$ using mid-infrared data. The physical reason for the role of emission at wavelength $\gtrsim 10000$ \AA\ in determining $M_\ast$ also will be studied once more data is available.

\begin{acknowledgements}
This work is supported by the Strategic Priority Research Program of Chinese Academy of Sciences (Grant No. XDB 41000000), the National Science Foundation of China (NSFC, Grant No. 12233008, 11973038), the China Manned Space Project (No. CMS-CSST-2021-A07) and the Cyrus Chun Ying Tang Foundations.
Z.S.L. acknowledges the support from the China Postdoctoral Science Foundation (2021M700137). Y.Z.G. acknowledges support from the China Postdoctoral Science Foundation funded project (2020M681281).
\end{acknowledgements}

\bibliography{ref}
\bibliographystyle{aasjournal}

\end{document}